\documentclass[12pt]{article}
\usepackage{epsfig,epsf,rotating}
\begin{document}

\title{R\'enyi statistics in high energy particle production}

\author{A.Kropivnitskaya, A.Rostovtsev\\
\\
{\it Institute f. Theoretical and Experimental Physics, ITEP,} \\
{\it Moscow, Russia}
}
\date{ }
\maketitle
\begin{center}
Abstract
\end{center}
It is shown that R\'enyi statistics provides a plausible basis to 
describe the hadron distributions measured in high energy particle 
interactions.
Generalized Boltzmann and gamma distributions obtained by maximization of
R\'enyi entropy under constraints on Kolmogorov-Nagumo averages
are used to describe the hadron transverse momentum and multiplicity spectra
correspondingly.
\vspace*{1.0cm}

%\section{Introduction}
A multi-hadron production in high energy hadronic collisions is
a complicated phenomenon generally modeled by a dynamic
system of hadronizing quarks and gluons with many
degrees of freedom and high level of correlations.
Though using QCD one calculates elementary
interactions of quarks and gluons at microscopic level, a complexity
of the system generally
doesn't allow an extension of these calculations to predict
the observed particle spectra. Thus, it is tempting to apply
statistical methods to understand properties of particle
production presented by the experimentally measured statistical distributions.
In this paper we consider two distributions
most systematically studied in experiments with $pp$ and $\bar{p}p$
interactions:
the inclusive charged particle transverse momentum spectrum and
charged particle multiplicity distribution.

At low collision energy~$(s)$ an exponential behavior of the 
inclusive single particle
spectra as function of particle's transverse momentum~$(P_t)$
has been observed. These spectra are interpreted using a
thermodynamic analogy and described by a Boltzmann-type
distribution~\cite{Hagedorn}
\footnote{Since the most data on hadron
production are available for central rapidities~($y=0$)
we consider the differential cross sections for central
rapidity only.}
\begin{equation}
\left.\frac{d^3\sigma}{d^3p}\right|_{y=0} =
A\cdot e^{-E_t/T}\,,
\label{Boltzmann}
\end{equation}
where $A$ is a normalization factor, $E_t=\sqrt{m^2+P_t^2}$ is a 
transverse energy of the produced particle with mass $m$, and $T$ is a 
characteristic temperature of the hadronizing system. 
What is a genesis of the exponential distribution?

The least biased method to obtain statistical distributions,
which are realized in the nature
was promoted by E.T.Jaynes as Maximum Entropy 
Principle~(MEP)~\cite{Jeynes}.
The MEP states that the physical observable has
a distribution, consistent with given constraints which maximizes
the entropy. 
The Boltzmann exponential distribution~(\ref{Boltzmann})
arises naturally from a maximization of Gibbs-Shannon entropy 
\begin{equation}
S_{GS} = -\sum_{n} P_n ln(P_n)
\label{GS}
\end{equation}
under a constraint on an average energy of produced hadrons.

The data on high-$s$ interactions show a
deviation from the exponential towards a power-law behaviour
and can be approximated~\cite{Beck} by
\begin{equation}
\left.\frac{d^3\sigma}{d^3p}\right|_{y=0} =
A\cdot (1+\frac{E_t}{\kappa T})^{-\kappa}\,.
\label{Power-law}
\end{equation}
At finite values of 
the parameter $\kappa$ (\ref{Power-law}) represents a power-law 
distribution, while in the limit of $\kappa \rightarrow \infty$  
the expression~(\ref{Power-law})
reduces to Boltzmann exponent~(\ref{Boltzmann}).
In plasma physics the $\kappa$-distribution~(\ref{Power-law})
is frequently used to describe an excess of highly energetic 
particles with respect
to that expected from an exponential spectrum~\cite{Treumann}.
The enhanced probability of the high energy particle fluctuations (or an 
appearance of a {\it heavy tail} in a distribution)
could be viewed as a collective phenomenon resulting from a strong
intrinsic correlation built in the system.
The heavy-tailed distribution~(\ref{Power-law}) arises in a framework of
non-extensive statistical mechanics~\cite{Tsallis} as a result of
maximization of Harvda-Charvat-Dar\'oczy-Tsallis~(HCDT) 
 entropy~\cite{HCDT} 
\begin{equation}
S_q = \frac{1 - \sum P_n^q}{q-1}\,.
\label{HCDT}
\end{equation}
under a constraint on an average energy of the particles.
Formally, the expression~(\ref{Power-law}) with $\kappa=(q-1)^{-1}$ is a 
non-extensive generalization of Boltzmann distribution.

  It is interesting to note that the exponential 
distribution~(\ref{Boltzmann}) transforms in to~(\ref{Power-law}) when
a value of parameter $T$ fluctuates according to a gamma distribution~\cite{Wilk}
\begin{equation}
f(T,\lambda,k) = \frac{\lambda(\lambda T)^\mu \cdot e^{-\lambda T}}
{\Gamma(k)}\,\,.
\label{Gamma}
\end{equation}
The gamma distribution represents a sum of $\mu+1$ 
independent exponentially distributed random 
variables and arises naturally from maximization of Gibbs-Shannon 
entropy under constraints on the mean values of $T$ and $\ln(T)$. The latter 
requirement corresponds to a constraint on a geometric mean. 

There are a number of similarities between a behaviour of hadron $P_t$ spectra
and hadron multiplicity distributions. The multiplicity distributions are observed
to be significantly broader ({\it over-dispersed}) than 
the Poisson distribution expected for uncorrelated systems. The 
data on 
charged hadron multiplicities are frequently parameterized by the Negative 
Binomial Distribution~(NBD)~\cite{Eddi}. 
The NBD arises from Poisson
distribution with the mean value fluctuating according to a gamma distribution.
In other words, the NBD results from a discrete gamma distribution with each point smeared by a 
correspondent Poisson distribution.
Like the heavy tail in hadron $P_t$ spectra, 
an over-dispersion of hadron multiplicity distribution is believed to be a 
result of strong long-range correlations in the system.
It is therefore tempting to find a common statistical 
approach in which both the heavy-tailed and over-dispersed distributions 
arise naturally. 
Below it is shown that R\'enyi statistics with 
Kolmogorov-Nagumo averages provides a plausible basis to derive such
statistical distributions. 
\vspace*{1.0cm}

%\section{Boltzmann distribution in R\'enyi statistics and hadron $P_t$ 
%spectra}
The generalized Kolmogorov-Nagumo~(KN) average of a variable $x_n$ is 
defined as
\begin{equation}
<x> = \phi^{-1}(\sum_{n} P_n\phi(x_n))\,,
\label{KN}
\end{equation}
where $\phi$ is a monotonic function and $P_n$ is a probability of $n$-th 
state. For $x_n=\ln(P_n)$ the average~(\ref{KN}) represents a generalized 
measure of information (entropy). If one requires the generalized entropy 
to be additive (extensive) and $<x+c>=<x>+c$, where $c$ is a constant, 
only linear and exponential functions $\phi(x)$ satisfy these requirements.
It is practical to represent the function $\phi(x)$ in a 
generalized form
\begin{equation}
\phi(x) = \frac{e^{(1-q)x} - 1}{1-q} \,,
\label{phi}
\end{equation}
which reduces to a linear function in the limit $q\rightarrow 1$.
The linear function leads to the traditional definition of an average and 
to Gibbs-Shannon entropy~(\ref{GS}), while the generalized 
function~(\ref{phi}) leads to a definition of R\'enyi entropy~\cite{Renyi}
\begin{equation}
S_q = \frac{ln(\sum P_n^q)}{1-q}\,.
\label{Renyi}
\end{equation}
For $|q-1|\ll 1$ the $S_q$ can be 
approximated by non-extensive HCDT entropy~(\ref{HCDT}), while for $q\rightarrow 1$ the 
entropy~(\ref{Renyi}) reduces to 
Gibbs-Shannon form~(\ref{GS}). 
The Gibbs-Shannon statistics has no built-in
intrinsic correlations, such that for two variables $x$ and $y$
holds $<x+y>=<x>+<y>$, which is not 
necesseraly a case for the R\'enyi statistics. 
Although R\'enyi entropy provides the most general form of information measure,
it has found so far a limited number of applications. 
One of the reasons for that is
a little understanding of physical meaning of the entropic index~$q$.
Since for $q=1$ the  Gibbs-Shannon statistics is restored it is reasonable to
assume that a deviation of the entropic index value from unit is related
to intrinsic correlations built up in a system. As we will see below,
this intuitive conjecture is compatible with the hadronic interaction data.

In order to obtain a generalized Boltzmann distribution arising in the R\'enyi 
statistics one has to maximize $S_q$ under a constraint on
a KN-average value. This was done 
in~\cite{Marek} and in application to a  hadron $P_t$ spectrum can be
rewritten as
\begin{equation}
\left.\frac{d^3\sigma}{d^3p}\right|_{y=0}
 = A\cdot
(1-\beta + \beta\cdot e^{\frac{(q-1)\lambda E_t}{\beta}})^{-\frac{1}{q-1}}\,,
\,\,\,\,\,\,\, q > 1
\label{B1}
\end{equation}
with a parameter $\lambda=(q T)^{-1}$ needed to make the energy 
dimensionless and $\beta$ being Lagrange multiplier.
For $|(q-1)\lambda E_t/\beta|\ll 1$ the 
expression~(\ref{B1}) is approximated by the power-law 
distribution~(\ref{Power-law}) and the Boltzmann exponential 
distribution~(\ref{Boltzmann}) follows in the limit 
$q\rightarrow 1$. 

\begin{figure}[h]
\begin{center}
%\hspace*{-5.0cm}
\epsfig{
file=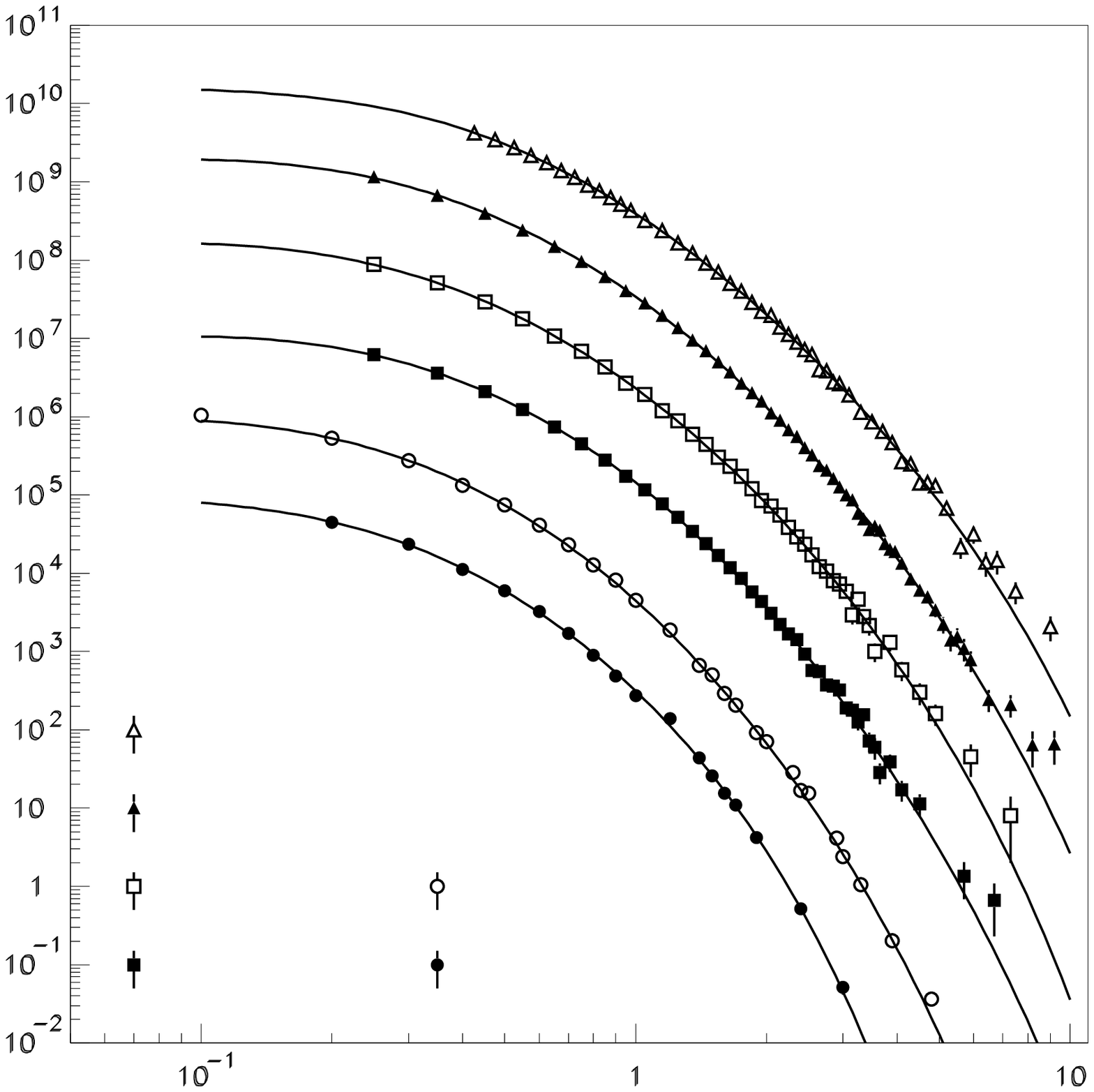,
%        bburx=500,
%        bbury=543,bbllx=69,bblly=0,
        height=15.0cm,width=14.0cm,angle=0} 
\put(-300,170){\large\bf $\sqrt{s}~[GeV]$ } 
\put(-340,170){\large\bf $\bar{p}p$ } 
\put(-250,116){\large\bf $pp$}
\put(-320,142){\large\bf $1800$}
\put(-320,116){\large\bf $900$}
\put(-320,90){\large\bf $500$}
\put(-320,64){\large\bf $200$}
\put(-225,90){\large\bf $53$}
\put(-225,60){\large\bf $23$}
\put(-352,248){\small\bf $\times 10$}
\put(-352,276){\small\bf $\times 10^2$}
\put(-352,307){\small\bf $\times 10^3$}
\put(-352,336){\small\bf $\times 10^4$}
\put(-352,360){\small\bf $\times 10^5$}
\put(-120,10){\large\bf $P_t~[GeV]$}
\put(-420,220){\begin{sideways}\Large\bf
$E\left.\frac{d^3\sigma}{d^3P}\right|_{y=0} [\mu{b}/GeV^2]$\end{sideways}}
\caption{Measured charged particle $P_t$ spectra overlayed with
fits to the generalized Boltzmann distribution.}
  \label{Pt}
\end{center}
\end{figure}

\begin{figure}[h]
\begin{center}
%\hspace*{-5.0cm}
\epsfig{
file=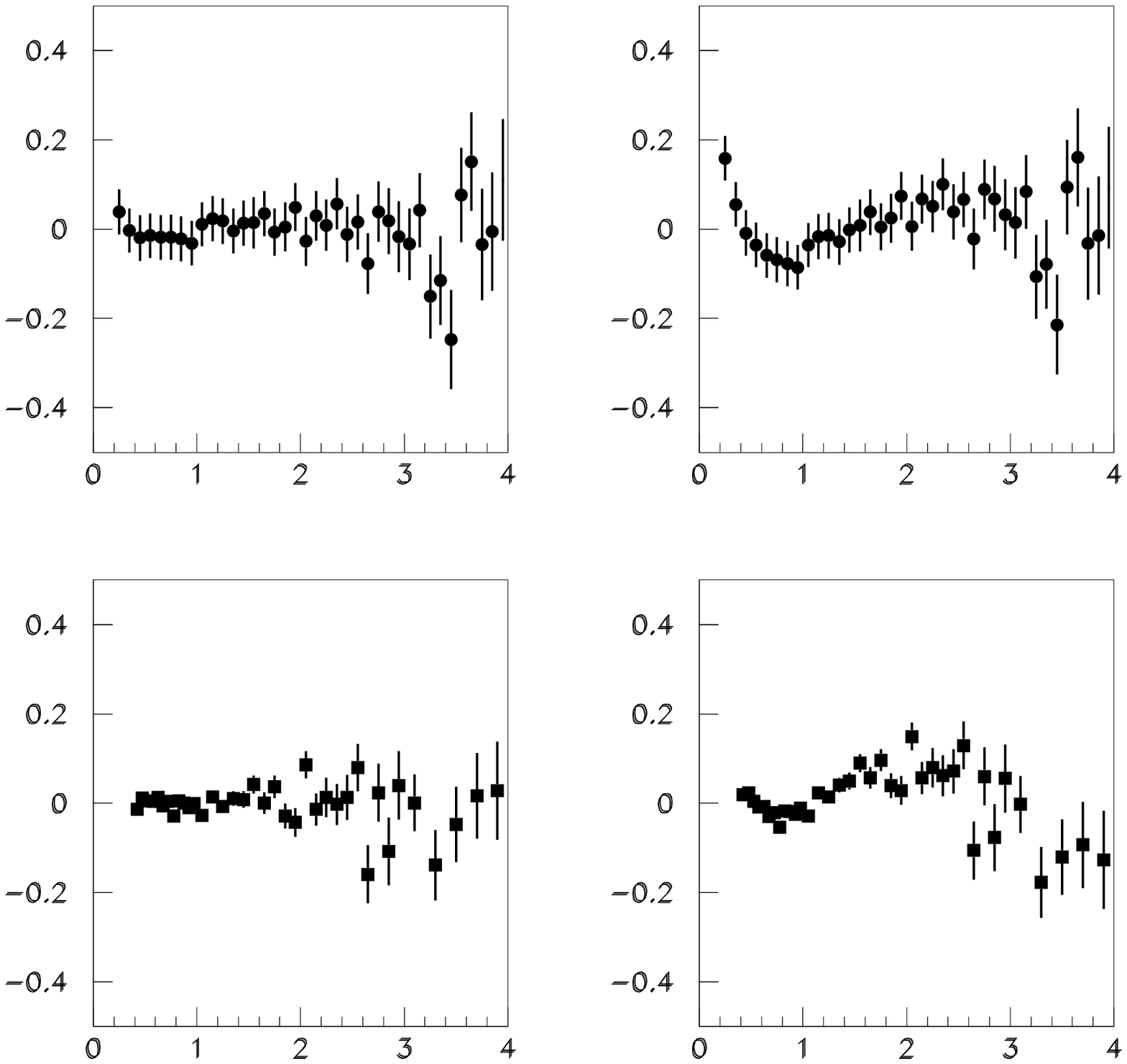,
%        bburx=500,
%        bbury=543,bbllx=69,bblly=0,
        height=15.0cm,width=14.0cm,angle=0}
\put(-255,360){\Large\bf a}
\put(-70,360){\Large\bf c}
\put(-255,160){\Large\bf b}
\put(-70,160){\Large\bf d}
\put(-340,250){\large\bf $\sqrt{s}=1800GeV$ } 
\put(-340,60){\large\bf $\sqrt{s}=900~GeV$}
\put(-150,250){\large\bf $\sqrt{s}=1800GeV$ } 
\put(-150,60){\large\bf $\sqrt{s}=900~GeV$}
\put(-120,10){\large\bf $P_t~[GeV]$}
\put(-420,160){\begin{sideways}\Large\bf
(Data - Fit)/Fit\end{sideways}}
\caption{A relative difference between the measured 
charged particle $P_t$ spectra and 
the fit results using the generalized Boltzmann~(a and b) 
and the power-law HCDT parameterization~(c and d).}
  \label{sub}
\end{center}
\end{figure}

In Fig.~\ref{Pt} the experimental spectra measured at different $pp$ and $\bar{p}p$ 
collision energies are shown together with
the parameterization~(\ref{B1}). The 
experimental data come from~\cite{Pt-data}. 
In order to make a comparison with measured differential charged 
particle cross section 
$Ed^3\sigma/d^3p \equiv d^2\sigma/(2\pi P_tdP_tdy)$ at $y=0$ 
the right part of~(\ref{B1}) is multiplied by $E_t$ and
$m$ in $E_t$ fixed to charged pion mass value.
For all data sets a good data description in Fig.~\ref{Pt} is observed. To 
compare a quality of the data description using the 
power-law~(\ref{Power-law}) and the generalized 
Boltzmann~(\ref{B1}) distributions the residuals
of the correspondent fits to the data are shown in Fig.~\ref{sub} for two highest 
available collision energies. It is clear that the
power-law function has local systematic deficiencies, while the  
generalized Boltzmann distribution describes the data equally well in the 
whole available range of the transverse momenta.

%\section{Normal distribution in R\'enyi statistics and hadron 
%multiplicity spectra}
In order to derive a generalized gamma distribution one has to
maximize entropy $S_q$
under constraints on KN-mean values of $x_n$ and $\ln(x_n)$ 
\begin{equation}
\frac{\ln(\sum P_n e^{\frac{\lambda x_n (1-q)}{\beta}} -1)}{1-q} \equiv <x>\,,
\,\,\,\,\,\,\,\,
\frac{\ln((\sum P_n x_n^{\mu(1-q)}) -1)}{1-q} \equiv <ln(x)>\,.
\label{constr}
\end{equation}
In our application, $x_n=N$ is the number of produced charged hadrons and the resulting 
distribution is read 
\begin{equation}
 P_N 
 = A\cdot
((\lambda N)^{-\mu (q-1)} 
-\beta + \beta e^{\frac{(q-1)\lambda N}{\beta}})^{-\frac{1}{q-1}}\,,
\,\,\,\,\,\,\, q > 1\,,
\label{B2}
\end{equation}
For $q\rightarrow 1$ the~(\ref{B2}) is reduced to the 
gamma distribution~(\ref{Gamma}).

\begin{figure}[h]
\begin{center}
%\hspace*{-5.0cm}
\epsfig{
file=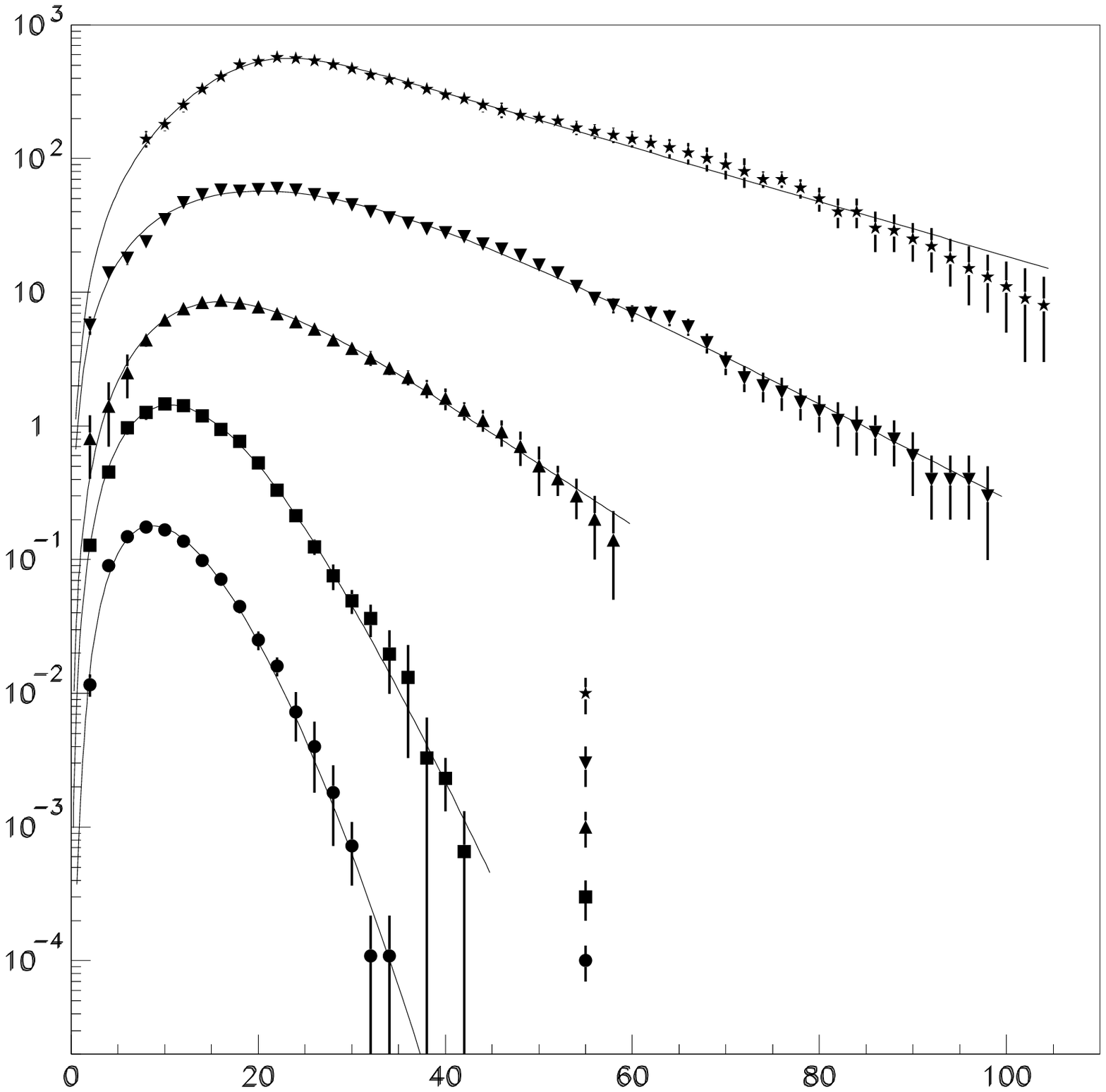,
%        bburx=500,
%        bbury=543,bbllx=69,bblly=0,
        height=15.0cm,width=14.0cm,angle=0}
\put(-150,185){\large\bf $\sqrt{s}~[GeV]$}
\put(-180,157){\large\bf $900$}
\put(-180,135){\large\bf $540$}
\put(-180,113){\large\bf $200$}
\put(-180,89){\large\bf $53$}
\put(-180,70){\large\bf $30$}
\put(-250,170){\large\bf $\times 10$}
\put(-220,255){\large\bf $\times 10^2$}
\put(-190,300){\large\bf $\times 10^3$}
\put(-160,350){\large\bf $\times 10^4$}
\put(-100,10){\large\bf $N$}
\put(-405,355){\large\bf $P_N$}
\caption{
Measured charged particle multiplicity spectra overlayed
with fits to the generalized gamma distribution.}
  \label{N-fit}
\end{center}
\end{figure}

In Fig.~\ref{N-fit} the experimental charged multiplicity spectra 
measured at different collision energies~\cite{N-data} are fit to the 
parameterization~(\ref{B2}). For all data sets
a good description by the generalized gamma 
distribution~(\ref{B2}) is found.
The fits to the data show that the parameter $q$ increases logarithmically with~$s$ 
both for $P_t$ and multiplicity spectra. At the same time the experiments show
a similar logarithmic rise of a particle-particle correlation strength with~$s$~\cite{corr}.
This observation provides a support for the conjecture that 
the entropic index $q$ is related to the intrinsic correlations in system.
Note, that for $\mu = 0$ the expression~(\ref{B2}) transforms in to
the generalized Boltzmann exponential distribution~(\ref{B1}).

In conclusion, using the Maximum Entropy Principle two generalized statistical distributions have 
been derived from a maximization of R\'enyi entropy. These distributions have been applied to 
describe the experimental data on particle production in high energy $pp$ and $\bar{p}p$ 
interactions. The generalized Boltzman distribution turned
out to describe the $P_t$ spectra of produced charged particles better than the traditionally used 
power-law distribution. The generalized gamma distribution provides a good description of
the charged particle multiplicity.

%%%%%%
%%%%%%

\section*{Acknowledgments}

The authors gratefully acknowledge 
P.Harremoes, A.Paramonov, R.A.Treu- mann and C.Tsallis for 
suggestive discussions.
The work was partially supported by
Russian Foundation for Basic Research grants RFBR-01-02-16431
and RFBR-03-02-17291.

\end{document}